# The Effect of Ionic Composition on Acoustic Phonon Speeds in Hybrid Perovskites from Brillouin Spectroscopy and Density Functional Theory

Irina V. Kabakova,[1,2,*] Ido Azuri,[3] Zhuoying Chen,[4] Pabitra K. Nayak,[5] Henry J. Snaith,[5] Leeor Kronik,[3] Carl Paterson,[1] Artem A. Bakulin,[6] and David A. Egger[7,*]

[1] *Blackett Laboratory, Imperial College London, Kensington, London SW7 2AZ, UK*

[2] *School of Mathematical and Physical Sciences, University Technology Sydney, NSW 2007, Australia*

[3] *Department of Materials and Interfaces, Weizmann Institute of Science, Rehovoth 76100, Israel*

[4] *Laboratoire de Physique et d'Etude des Matériaux (LPEM), CNRS-UMR 8213, ESPCI Paris, Université Pierre et Marie Curie, 75005 Paris, France*

[5] *Clarendon Laboratory, Department of Physics, University of Oxford, Parks Road, Oxford, OX1 3PU, UK*

[6] *Department of Chemistry, Imperial College London, Kensington, London SW7 2AZ, UK*

[7] *Institute of Theoretical Physics, University of Regensburg, 93040 Regensburg, Germany*

* Correspondence: irina.kabakova@uts.edu.au; david.egger@physik.uni-regensburg.de

**Abstract**

Hybrid organic-inorganic perovskites (HOIPs) have recently emerged as highly promising solution-processable materials for photovoltaic (PV) and other optoelectronic devices. HOIPs represent a broad family of materials with properties highly tuneable by the ions that make up the perovskite structure as well as their multiple combinations. Interestingly, recent high-efficiency PV devices using HOIPs with substantially improved long-term stability have used combinations of different ionic compositions. The structural dynamics of these systems are unique for semiconducting materials and are currently argued to be central to HOIPs stability and charge-transport properties. Here, we studied the impact of ionic composition on phonon speeds of HOIPs from Brillouin spectroscopy experiments and density functional theory calculations for $FAPbBr_3$, $MAPbBr_3$, $MAPbCl_3$, and the mixed halide $MAPbBr_{1.25}Cl_{1.75}$. Our results show that the acoustic phonon speeds can be strongly modified by ionic composition, which we explain by analysing the lead-halide sublattice in detail. The vibrational properties of HOIPs are therefore tuneable by using targeted ionic compositions in the perovskite structure. This tuning can be rationalized with non-trivial effects, for example, considering the influence of the shape and dipole moment of organic cations. This has an important implication to further improvements in the stability and charge-transport properties of these systems.

Hybrid organic-inorganic perovskites (HOIPs) have recently emerged as highly promising materials for optoelectronic applications, especially for high-efficiency photovoltaic (PV) and light-emitting devices; see refs. [1–7] for reviews. HOIPs are crystalline systems with $ABX_3$ stoichiometry: typically, inorganic ions occupy the $B$ and $X$ sites at corner-sharing octahedra, and organic cations the $A$ sites, *i.e.*, are at octahedral voids.[8] Among the many outstanding properties of these crystals, we highlight their ionic compositional flexibility that, within the geometrical constraints of the perovskite structure,[9] provides means to tune HOIP electronic and optical properties. Substituting different halogen ions (halides, *e.g.*, I, Br, Cl) at the $X$ site allows for tuning of the band gap of HOIPs.[10,11] Furthermore, using I and Cl to prepare a mixed HOIP was shown to modify strongly the charge-carrier lifetime,[12] a fundamentally important parameter for PV devices. Interestingly, today's best HOIP PV cells mix different cations at the $A$ site of HOIPs, *i.e.*, organic (*e.g.*, methylammonium, MA or formamidinium, FA) and inorganic ones (*e.g.*, Cs, K or Rb).[13–16] It is therefore of fundamental and practical interest to understand microscopically the effect of $A$ and $X$ site substitution on HOIP properties.

HOIPs are mechanically soft materials with rather low elastic constants compared to other semiconductors used in efficient solar-cell devices.[17–20] This is expected, because the rather small ionic charges in the halide perovskite structure result in a weaker electrostatic interaction of the lattice ions.[21] The mechanically soft nature of HOIPs appears to be especially relevant for the material stability of HOIPs, one of the major practical issues of these materials. Some HOIPs are known to be water soluble and may even be partially unstable enthalpically;[22] they can also exhibit ion migration[23–25] and phase segregation[26,27] at room temperature. Indeed, different ionic compositions have been shown to affect these properties strongly, with reported impacts on material stability,[16,28,29] water tolerance,[30] and ion diffusion.[31] Moreover, a soft lattice implies low phonon speeds and the existence of low energy phonons, which may have direct implications for transport phenomena in HOIP-based devices: lower acoustic phonon speeds imply larger acoustic-phonon scattering rates and lower mobilities; the phonon speed was also shown to determine the thermal transport properties of HOIPs.[32] To improve the stability and device performance of HOIP PV cells further, it is important to understand how, and to what extent, their mechanical properties can be modified by ionic composition.

In this Letter, we report the phonon speeds of HOIPs with different ionic compositions using a combined experimental and theoretical approach. The effect of *A* and *X* site substitution was studied with Brillouin spectroscopy (BS) and density functional theory (DFT) calculations for the HOIPs FAPbBr$_3$, MAPbBr$_3$, MAPbCl$_3$, and the mixed halide MAPbBr$_{1.25}$Cl$_{1.75}$. Our data confirm the by now well-known mechanically soft nature of the lattice for a broader range of HOIP systems, and are in line with recent results[33] on MAPbBr$_3$, using similar experimental techniques. Here, we go significantly beyond in showing that the acoustic phonon speeds and elastic properties strongly depend on the specific ionic composition. By combining BS with theoretical calculations for different *A* and *X* site substituted HOIPs, we explain this effect by analysing the lead-halide sublattice and highlight the contribution of non-trivial effects, such as the influence of the shape and dipole moment of organic cations. Our findings provide insights to improve the mechanical stability and tune structurally dynamic properties of HOIPs by targeted ionic composition, which has direct implications for PV device performance.

BS probes acoustic vibrations at GHz frequencies (~0.1-1 cm$^{-1}$) and can be used to measure material elasticity.[34] An incoming photon (angular frequency $\omega_i$, wave vector $k_i$) interacts with a phonon (frequency $\Omega$, wave vector K), and is scattered at an angle $\Theta$ between $k_i$ and $k_s$, the wave vector of the scattered photon (see Fig. 1a). The frequency of the scattered photon, $\omega_s$, is shifted by the phonon frequency, *i.e.*, $\omega_s=\omega_i\pm\Omega$; the ($\pm$) sign accounts for a phonon being created or annihilated in the scattering process. From momentum conservation one can obtain the frequency shift for a given mode, given by $\Omega_i = \frac{2nv_i}{\lambda}\sin(\frac{\Theta}{2})$, where n is the refractive index of the scattering medium, $\lambda$ the vacuum wavelength of the probing light, and $v_i$ the speed of the mode. In this work we only examined scattering angles of $\Theta = 180°$ or used a backscattering geometry setup (see Figs. 1b & 1c), for which $\Omega_i = \frac{2nv_i}{\lambda}$. See the experimental methods section for further details.

The measurements have been performed on single-halide and mixed-halide perovskite single crystals with a typical size of ~0.5 mm, see methods section for preparation details. The crystals had well-recognizable cubic shape and for the measurements were placed on the horizontal sample holder without additional alignment. Sample composition was tested using X-ray diffraction (XRD) spectroscopy. To determine the exact composition of mixed halide crystals, only XRD data were used (see supplementary information, SI) because self-absorption had a strong effect on the sample's absorption and emission.[35]

Fig. 2 shows the BS spectra for the different pure and mixed HOIPs, in which each acoustic mode is characterized by a unique frequency shift, $\Omega_i$, and, hence, a propagation speed, $v_i$. For BS in an anisotropic medium, the direction of light propagation determines the set of modes relevant for the scattering event: propagation parallel to a high symmetry crystal axis involves pure longitudinal and transverse acoustic waves. For any other propagation direction, one of the involved vibrations is purely transverse and the other two are quasi-longitudinal and quasi-transverse modes.[34] The peaks at 0 and 30 GHz are two interference orders at the probe wavelength and are due to elastic Rayleigh scattering events (denoted as R in Fig. 2a) and any stray laser light that may be present in the measurement. Peaks in the spectral range between 0 and 30 GHz are assigned to quasi-longitudinal and quasi-transverse Stokes and anti-Stokes acoustic waves, respectively, labelled by $L_{S/AS}$ and $T_{S/AS}$ in Fig. 2a. In our experiments the light-propagation direction is unknown *a priori*. However, turning the $MAPbCl_3$ crystal by 90° resulted in virtually identical BS frequencies (to within ~100-200 MHz, see SI). This indicates that the relevant acoustic mode propagation is probed in a direction that is close to being parallel to a principal crystal axis, which is important for our computational approach (*vide infra*). Fig. 2b shows that the ionic composition greatly influences the BS frequency shift, which increases along the sequence $FAPbBr_3$ → $MAPbBr_3$ → $MAPbBr_{1.25}Cl_{1.75}$ → $MAPbCl_3$. This finding implies that the phonon speed can be tuned *via* the ionic composition at both the *A* and *X* site of HOIPs.

To quantify this effect and obtain the peak position ($\Omega_i$), we first analysed the BS spectrum using a harmonic oscillator model with a Lorentzian line-shape function and the refractive indices of the materials (see experimental method section). The results of this analysis, summarized in Table 1, show that the highest frequency shift and acoustic speed occurs for quasi-longitudinal waves propagating in $MAPbCl_3$ crystals ($v_L$ = 4000 m/s). The acoustic speed of quasi-longitudinal modes is 3880 m/s in $MAPbBr_{1.25}Cl_{1.75}$ and 3620 m/s in MAPbBr$_3$, which shows the significant effect of *X* site substitution on the phonon velocity. Interestingly, *A* site substitution also affects the speed of quasi-longitudinal modes, it is 3380 m/s in FAPbBr$_3$, which is the lowest one found in all materials studied. For all HOIPs studied, quasi-longitudinal modes have a larger frequency shift than quasi-transverse ones. For MAPbCl$_3$ and FAPbBr$_3$, mechanically the most rigid and softest respectively of all materials considered, we could analyse the experimental data for quasi-transverse acoustic modes. We found that their speed is 45% of that of the quasi-longitudinal ones ($v_T$ = 1790 m/s) for the case of MAPbCl$_3$ and 62% ($v_T$ = 2110 m/s) for FAPbBr$_3$. Unfortunately, BS peaks associated with quasi-transverse acoustic waves in the other crystals, although indicating a similar trend, were too small to determine $\Omega_T$ reliably.

To study the effect of ionic composition on acoustic phonon speeds of HOIPs computationally, we performed DFT calculations, using the PBE functional[36] augmented by dispersive corrections,[37] as described previously.[38] To obtain acoustic phonon speeds, we used Christoffel's equation (see the theoretical methods section for further details). In HOIPs, the polar organic molecule rotates at room temperature exhibiting cubic symmetry on average, which we cannot take into account in our zero temperature DFT calculations. Rather, we assumed the organic molecule to be at a fixed orientation in the computational unit-cell, which means that cubic symmetry is broken and phonon speeds will be different along the high-symmetry axes in our calculations, see Table 1. As discussed above, the exact direction of probed phonon modes is unknown experimentally, and therefore we have to assume a propagation direction in our computational analysis. Because the experimental results indicate that propagation occurs close to parallel to a principal crystal axis, we computed phonon speeds along the principal crystal axes (see Table 1), and found that the (100) direction showed by far the best agreement between experiment and theory.

Fig. 3 shows a comparison of the theoretical data for $v_L$ along the (100) direction and values averaged over the principal crystal axes with the experimental results, as a function of Cl concentration in MAPbBr$_{3-x}$Cl$_{3-x}$. Both the (100) and averaged calculated data clearly show the large effect of *X* site substitution on acoustic phonon speeds, in agreement with the experimental trend. At a quantitative level we find very good agreement between the (100) theoretical data for MAPbBr$_3$ (2% deviation) and the mixed Br/Cl crystal (1% deviation), and a somewhat larger deviation of 5% for MAPbCl$_3$. Given the caveats mentioned above, perfect agreement between experiment and theory is not expected, even in principle. Nevertheless, the comparison in Fig. 3 shows that our computational approach can capture the physical effect of ionic composition on acoustic phonon speeds. We also note that our calculated data are in good agreement with previous theoretical results on HOIPs.[18,39]

For FAPbBr$_3$, our calculations yield a value of 3482 m/s along the (100) direction, again in reasonable agreement with the experimental value, although the effect of *A* site substitution is somewhat smaller as compared to the experimental result. The effect of *X* site substitution on acoustic phonon speeds can be well explained by the ionicity of the crystal, where larger electronegativity differences are expected to result in stronger ionic bonds. Hence, going from Br to Cl perovskites *via* the mixed Br/Cl crystal results in shorter bonds, smaller lattice constants and unit cells, larger elastic constants, and, thus, faster acoustic modes. However, the effect of *A* site substitution on acoustic phonon speed deserves further scrutiny because exchanging MA by FA hardly modifies the lattice constant as compared to halide exhange,[40] and we therefore studied the interaction of the organic molecule with the inorganic lattice in more detail.

While the electronic states of the organic cation are not close to the frontier states in the band structure, the interaction among the inorganic ions may still be affected electrostatically by the presence of the molecule. Therefore, Fig. 4 shows the DFT-computed electrostatic potential energy at an *xz*-plane placed at a *y*-position close to a nitrogen atom in MAPbBr$_3$ (panel a) and FAPbBr$_3$ (panel b). Comparing the two cases, it can be seen that the effect of the molecule is different: the spherical symmetry of the potential energy close to the proximal Br atoms is preserved in MAPbBr$_3$, but strongly distorted in FAPbBr$_3$. Furthermore, the gradient in potential energy, which can be gleaned from the isolines in Fig. 4, is more pronounced and confined in MAPbBr$_3$ than in FAPbBr$_3$. Therefore, in FAPbBr$_3$ the presence of the molecule affects the charge density in its vicinity, which impacts the inorganic lattice such that it is less ionic and weaker electrostatically. At first sight, this result may seem counterintuitive, as the molecular structure of MA appears to be less symmetric than that of FA. However, the ionic size of FA is larger than that of MA,[16] which means that the mutual interaction between FA and the inorganic surrounding is spatially more extended. Moreover, the charge density of MA is more confined around the nitrogen, whereas it is much more delocalized in FA, and therefore the dipole moment of FA is lower by a factor of ten compared to MA.[21,41] As a result, for FA the Madelung energy arising from the presence of the dipolar molecule is smaller, as are the charges on the inorganic ions in $FAPbBr_3$ compared to $MAPbBr_3$. Therefore, the size of FA and its small dipole moment weaken the bonding of the inorganic lattice, which reduces elastic constants and acoustic phonon speeds.

In summary, we studied the impact of ionic composition on phonon speeds of HOIPs by combining Brillouin spectroscopy experiments and density functional theory calculations for $FAPbBr_3$, $MAPbBr_3$, $MAPbCl_3$, and the mixed halide $MAPbBr_{1.25}Cl_{1.75}$. Our findings show that the acoustic phonon speeds can be strongly modified by both *A* and *X* site substitution, which we explained by considering the iconicity, mechanical strength, and lattice constants of the lead-halide sublattice. Interestingly, our theoretical results imply that *A* site substitution affects lead-halide bonding electrostatically, which means that the vibrational properties of HOIPs may be tuned by combining cations with specific dipole moments and sizes. Because lower acoustic phonon speeds imply stronger acoustic-phonon scattering of charge carriers and lower mobilities, in addition to smaller elastic constants, semiconductors with faster acoustic phonon speeds appear more suitable for devices. Such insight is important to further enhance the mechanical stability, as well as charge-transport characteristics of HOIPs, by targeted ionic composition.

**Methods section**

*Sample preparation*

*Materials.* Methylamine ($CH_3NH_2$, 40 wt% in water), hydrochloric acid (HCl, 37 wt% in water), diethyl ether (ACS reagent, ≥99.8%), *N-N*, dimethylformamide (DMF, anhydrous, 99.8%), lead (II) chloride ($PbCl_2$, 99.999%), and lead (II) bromide ($PbBr_2$, ≥98%) were purchased from Sigma Aldrich. Ethanol (anhydrous) was purchased from Carlo Erba Reagents, FABr from Dyesol, and hydrobromic acid (HBr) (47 wt% in water) and dimethylsulfoxide (DMSO) were purchased from Merck.

*$CH_3NH_3Cl$ (MACl) and $CH_3NH_3Br$ (MABr).* For MACl, HCl was added dropwise into $CH_3NH_2$ under stirring, in a flask kept in an ice bath according to the molar ratio $CH_3NH_2$:HCl = 1.2:1. This mixture was kept under stirring for 2 hours before it was evaporated at 55 °C in a rotary evaporator for solvent removal. The obtained powders were dissolved in anhydrous ethanol and precipitated, by adding diethyl ether and subsequent centrifugation and decantation. This purification procedure was repeated two more times. The obtained powders were dried in a vacuum oven at 60 °C overnight and then transferred into an Ar-filled glovebox for storage. MABr was synthesized by the same method as MACl, except that HBr was used instead of HCl.

*$CH_3NH_3PbCl_3$ (MAPbCl$_3$) crystals.* 2 mmol of $PbCl_2$ and 2 mmol of MACl was dissolved into 2 mL of 50% DMSO - 50% DMF mixture solution in a glass vial to obtain the concentration of 1 M. This glass vial was then kept undisturbed, in an oil bath at 55 °C, for 12 hours. The solvent was then discarded and the obtained MAPbCl$_3$ crystals were dried, under vacuum at 50 °C, for 6 hours.

*$CH_3NH_3PbBr_3$ (MAPbBr$_3$) crystals.* 4 mmol of $PbBr_2$ and 4 mmol of MABr was dissolved into 4 mL of DMF in a glass vial to obtain the concentration of 1 M. This glass vial was then kept undisturbed, in an oil bath at 100 °C, for 12 hours. The solvent was then discarded and the obtained MAPbBr$_3$ crystals were dried, under vacuum at 50 °C, for 6 hours.

*CH₃NH₃Pb(Br₁₋ₓClₓ)₃ (MAPb(Br₁₋ₓClₓ)₃) crystals.* MAPb(Br$_{1-x}$Cl$_x$)$_3$ crystals were synthesized according to the method, described by T. Zhang et al. in ref. [42]. Specifically, 2 mmol of PbBr$_2$, 1.1 mmol of MABr, and 0.9 mmol of MACl were dissolved into 2 mL of DMF in a glass vial to achieve a 1 M concentration of MAPb(Br$_{1-x}$Cl$_x$)$_3$ with x = 0.15. This glass vial was then kept undisturbed, in an oil bath at 52 °C, for 12 hours. The solvent was then discarded and the obtained mixed halide crystals were dried, under vacuum at 50 °C, for 6 hours. While x was 0.15 from the synthesis condition described above, the colour of the crystals suggested us that the obtained crystals may contain more Cl content. By analysing the position of XRD peaks in the pure- and mix-halide crystals (Table ST1) we then determined that the stoichiometric formula of the obtained crystals is MAPbBr$_{1.25}$Cl$_{1.75}$

*CH₅N₂PbBr₃ (FAPbBr₃) crystals*: FAPbBr$_3$ crystals were prepared by a slightly modified procedure, as reported for MAPbBr$_3$ in ref. [43]. Briefly, FABr and PbBr$_2$ salts were added to dimethyl formamide (DMF) to prepare a 2M solution. Formic acid (50 μl/ml) was added to the above solution followed by a filtration with a 0.45 micron filter. The filtered solution was poured into a glass vial. The glass vial was incubated at ~100°C to produce the FAPbBr$_3$ crystals. The crystals were collected when the solution is still hot (~ 100°C).

*Experimental methods*

Brillouin spectra of HOIPs were measured with the setup illustrated in Fig. 2. A 100 mW continuous-wave, frequency-doubled, fiber-coupled Nd:YVO$_4$ diode-pumped solid-state laser (CNI, MSL-FN-671) operating at λ=671 nm was used to probe the crystals. The spectrometer consisted of an interference filter, a virtual-image-phase-array (VIPA) etalon (Light Machinery) with a free spectral range of 30 GHz, and a CCD camera (Andor iXon DU888), as reported earlier.[44] The interference filter reduced the intensity of the elastic peak (scattered light not shifted in frequency) by 40dB, thereby improving the contrast ratio between elastic and Brillouin peaks in the scattering spectrum. The illumination light from the laser was collimated into an 8 mm diameter beam to fill the back focal plane of a 20x microscope objective (Olympus,

NA=0.5). Scattered light was collected by the same objective, separated from the incident light by the polarization beam-splitter and coupled into a single mode fibre, which also acted as a confocal pinhole. The quarter-wave plate inserted right before the objective was used to rotate the linear polarization axis of the return beam with respect to the illumination. The final imaging resolution of this setup was calculated to be 1x1x5 μm. The cubic-shaped perovskite crystals (FAPbBr$_3$, MAPbBr$_3$, MAPbCl$_3$, and MAPbBr$_{1.25}$Cl$_{1.75}$) were placed on the glass cover slip and fixed on the x-y-z moving stage. The average size of the crystals was approximately 1 mm$^3$, which was too small for the crystals to be polished. They were therefore measured as extracted from the growth chamber.

To determine the acoustic phonon speed we used refractive indices of $n_{Br}$=2.1 for all the bromide crystals and $n_{Cl}$=2 for the chloride crystal, as reported earlier.[18,33] The refractive index of the mixed crystal MAPbBr$_{1.25}$Cl$_{1.75}$ was estimated based on linear interpolation using $n_{BrCl} = xn_{Br} + (1-x)n_{Cl}$, where the fraction x=1.25/3≈0.42 was determined from X-ray diffraction, absorption, and photoluminescence tests.

*Computational methods*

We performed periodic DFT calculations using the VASP planewave code.[45] Exchange and correlation were treated at the level of the Perdew-Burke-Ernzerhof (PBE) functional.[36] PBE is well-known to not describe long-range dispersive interactions, which were shown to be important for computing accurate structural properties of HOIPs. We accounted for this by adding pair-wise dispersion interactions, computed using the planewave implementation[46,47] of the Tkatchenko-Scheffler (TS) pair-wise dispersion scheme,[37] to the PBE results. A convergence criterion of 10$^{-8}$ eV for the total energy and a planewave energy cutoff of 850 eV were used. To minimize residual forces and stresses, we used a threshold of 10$^{-3}$ eV/Å, with the exception of MAPbBr$_{1.25}$Cl$_{1.75}$, for which convergence was tedious and a force threshold of 8×10$^{-3}$ eV/Å was reached. For the optimization of the unit cell and atomic coordinates of FAPBr$_3$, we employed the GADGET tool based on internal coordinates.[48] Elastic constants were then calculated by perturbation theory, as implemented in VASP.[49] From these, we calculated the phonon speeds with Christoffel's equation for the general symmetry case,[50] and compared the results to those obtained with the ElIAM program.[51] 8x8x8 and 4x4x4 Γ-centred *k*-point grids have been used for calculations of the pseudo-cubic cells of all pure HOIPs and the 2x2x2 supercell of the mixed halide composition, respectively. Core-valence electron interactions were treated within the projector-augmented wave (PAW) formalism[52] using the 'normal' (*i.e.*, not 'soft'), program-supplied PAW potentials. We chose to treat the *5d* electrons of Pb explicitly. The effect of spin-orbit coupling has been neglected, as we have shown previously that these do not significantly affect the computation of structural parameters,[38] and hence can safely be assumed not to affect the elastic properties. The data shown in Fig. 4 were visualized using the XCrysDen program.[53]

**Acknowledgements**

We acknowledge support by a research grant from Dana and Yossie Hollander, in the framework of the Weizmann Institute of Science (WIS) sustainability and energy research initiative (SAERI). IVK acknowledges funding provided by Imperial College London through the ICRF program. ZC acknowledges the UPNAN project (in the framework of IDEX with the reference of ANR-10-IDEX-0001-02 PSL*) and the ANR-DFG project PROCES of the French National Research Agency (ANR). PKN acknowledges funding provided by Marie Skłodowska-Curie actions individual fellowships (grant agreement numbers 653184) and EPSRC, UK. AAB is a Royal Society University Research Fellow. DAE acknowledges funding provided by the Alexander von Humboldt Foundation in the framework of the Sofja Kovalevskaja Award endowed by the German Federal Ministry of Education and Research. The authors gratefully acknowledge the computing time granted by the John von Neumann Institute for Computing (NIC) and provided on the supercomputer JURECA[54] at Jülich Supercomputing Centre (JSC). We thank Dr. Zhiping Wang for support with the XRD measurements.

**Figure 1**

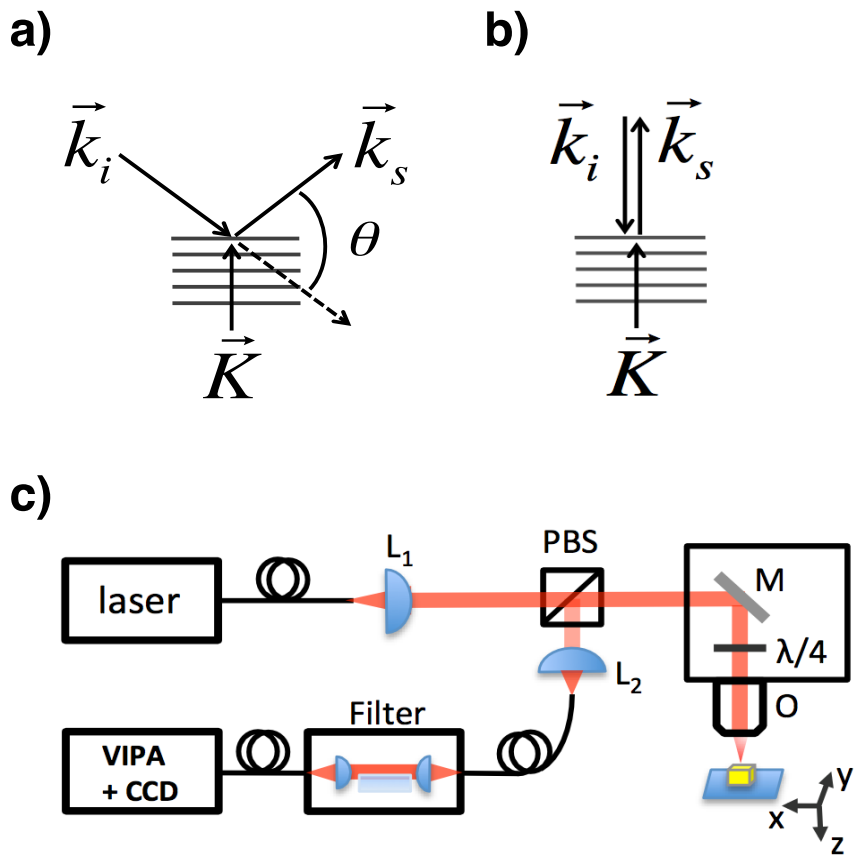

Fig. 1: Schematic of the inelastic Brillouin scattering process for an arbitrary scattering angle (a) and for a 180° scattering geometry (b); see text for details. (c) Schematic of the experimental setup for Brillouin spectroscopy measurements. $L_{1,2}$ are aspherical lenses, PBS a polarization beam-splitter, M a mirror, O is a microscope objective, and $\lambda/4$ is a quarter-wave plate. The spectrometer is based on a virtual-image-phase-array (VIPA) and a CCD camera.

**Figure 2**

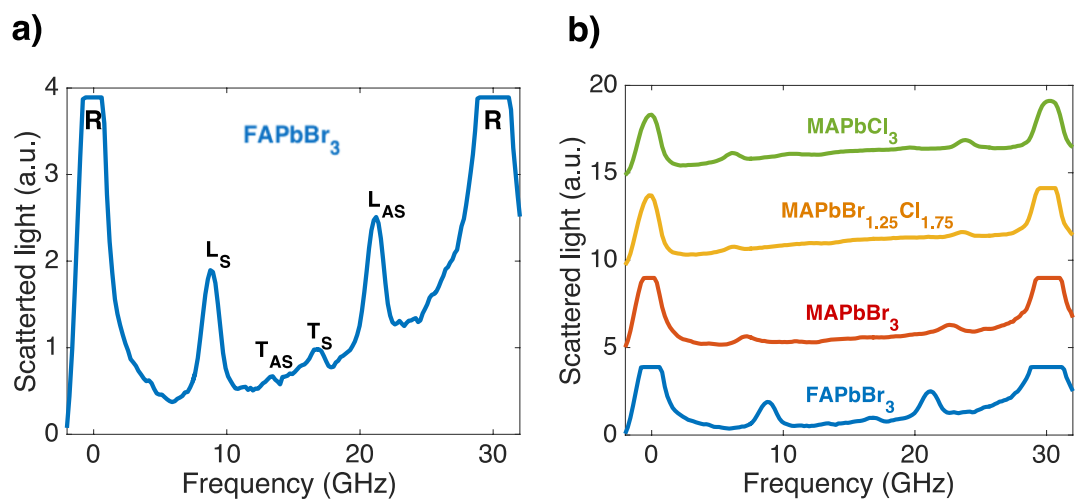

Fig. 2: a) Brillouin spectra, averaged across 50 individual measurements, for FAPbBr$_3$. R indicates Rayleigh peaks (two successive orders), L$_S$/L$_{AS}$ and T$_S$/T$_{AS}$ indicate the Stokes/anti-Stokes peaks associated with quasi-longitudinal and quasi-transverse acoustic waves, respectively. b) As a), but showing a comparison of FAPbBr$_3$, MAPbBr$_3$, MAPbBr$_{1.25}$Cl$_{1.75}$, and MAPbCl$_3$. Both graphs show one free spectral range of the VIPA spectrometer (30 GHz).

**Figure 3**

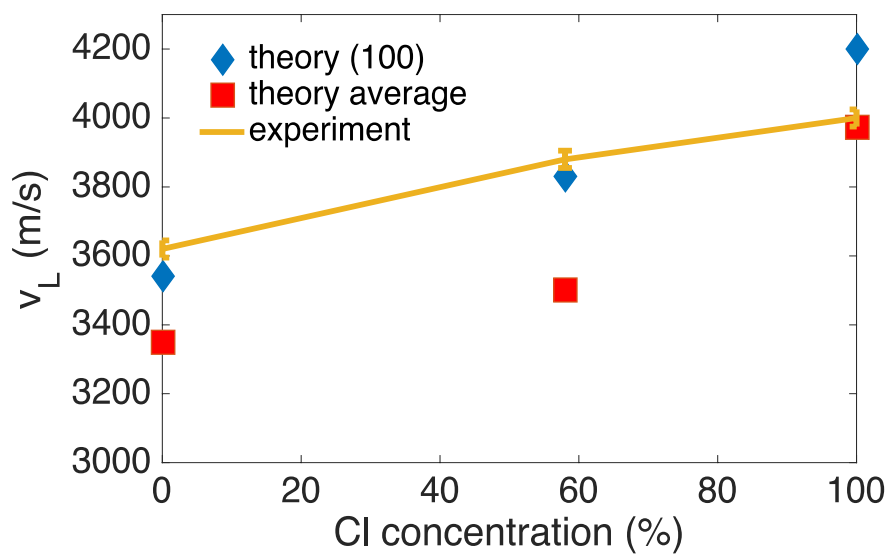

Fig. 3: Quasi-longitudinal speed of sound, $v_L$, measured experimentally (yellow line) and calculated theoretically as the average over the principal crystal axes (red diamonds) and along the (100) direction (blue squares) as a function of Cl concentration in MAPbBr$_{3-x}$Cl$_{3-x}$.

**Figure 4**

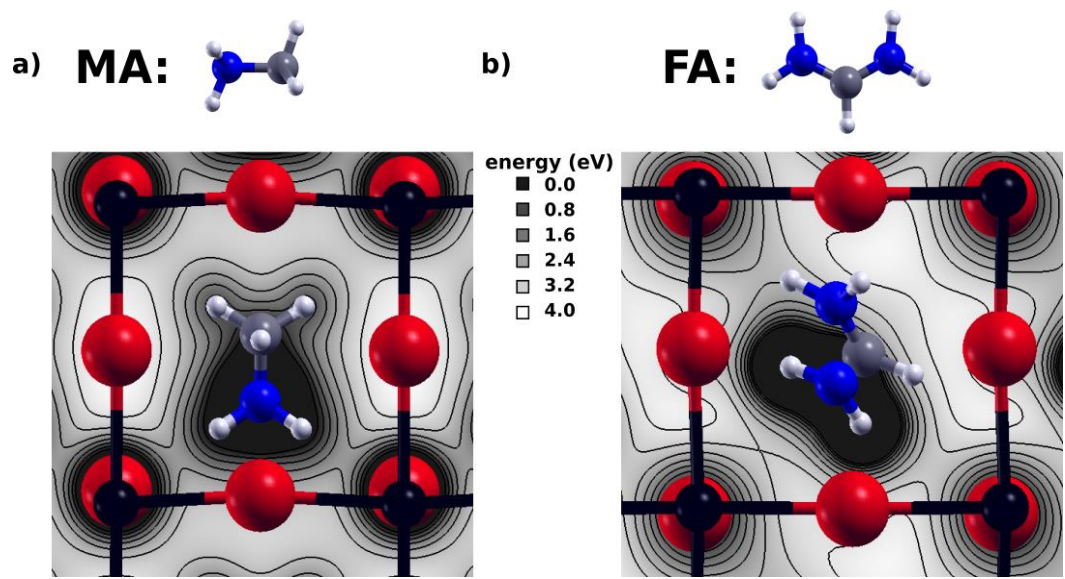

Fig. 4: DFT-computed electrostatic potential energy at an *xz*-plane, positioned 0.5 Å away from a nitrogen atom along the *y*-axis, for (a) MAPbBr$_3$ and (b) FAPbBr$_3$. The maximum of the energy scale corresponds to the maximum electrostatic energy per cell, and isolines are drawn every 0.5 eV. The figure shows lead (black), bromine (red), carbon (grey), nitrogen (blue), and hydrogen (white) atoms.

Table 1: Summary of the experimentally measured frequency shifts, Ω, and the measured and theoretically calculated acoustic phonons speeds, v, of different HOIPs for quasi-longitudinal (QL), quasi-transverse (QT), and transverse (T) modes.

| Crystal | Mode | Ω, GHz | Experiment v, m/s | Theory v, m/s [1 0 0] | [0 1 0] | [0 0 1] |
|---|---|---|---|---|---|---|
| MAPbCl$_3$ | QL | 23.9±0.1 | 4000±25 | 4201 | 3702 | 4020 |
|  | QT | 10.6±0.1 | 1770±25 | 1538 | 1396 | 1379 |
|  | T | − | − | 1205 | 1260 | 1255 |
| MAPbBr$_{1.25}$Cl$_{1.75}$ | QL | 23.6±0.1 | 3880±25 | 3832 | 3366 | 3308 |
|  | QT | − | − | 1391 | 1280 | 1230 |
|  | T | − | − | 1090 | 1162 | 1159 |
| MAPbBr$_3$ | QL | 22.7±0.1 | 3620±25 | 3540 | 3204 | 3306 |
|  | QT | − | − | 1266 | 1201 | 1169 |
|  | T | − | − | 1091 | 1096 | 1094 |
| FAPbBr$_3$ | QL | 21.2±0.1 | 3380±25 | 3482 | 3535 | 3529 |
|  | QT | 13.2±0.1 | 2110±25 | − | − | − |